\documentclass[10pt,twocolumn]{elsarticle}
\makeatletter
\def\ps@pprintTitle{%
  \let\@oddhead\@empty
  \let\@evenhead\@empty
  \let\@oddfoot\@empty
  \let\@evenfoot\@oddfoot
}
\makeatother
\usepackage{amssymb}
\usepackage{url}
\usepackage{xcolor}

\usepackage{hyperref}
\usepackage{amsmath,amsfonts}
\usepackage{graphicx}
\usepackage{textcomp}
\usepackage{adjustbox}
\usepackage{multirow}
\usepackage{threeparttable}
\journal{}

\begin{document}

\twocolumn[{\begin{frontmatter}

\title{Deep Learning methods for automatic evaluation of delayed enhancement-MRI. The results of the EMIDEC challenge.} %\tnoteref{tnote1}}%
%\tnotetext[tnote1]{This is an example for title footnote coding.}
{\footnotesize
\author[1,2]{Alain {Lalande}\fnref{fn1}\corref{cor1}}

\author[3]{Zhihao {Chen}\fnref{fn1}}
\author[4]{Thibaut {Pommier}}
\author[5]{Thomas~{Decourselle}}
\author[1]{Abdul {Qayyum}}
\author[3]{Michel {Salomon}}
\author[1]{Dominique~{Ginhac}}
\author[1]{Youssef {Skandarani}}
\author[1]{Arnaud {Boucher}}
\author[1,6,7]{Khawla {Brahim}}
\author[8,9,10]{Marleen {de Bruijne}}
\author[8,9]{Robin {Camarasa}}
\author[11,12]{Teresa~{M.~Correia}}
\author[13]{Xue {Feng}}
\author[1]{Kibrom B. {Girum}}
\author[14,15,16]{Anja {Hennemuth}}
\author[14,15]{Markus {Huellebrand}}
\author[1]{Raabid {Hussain}}
\author[14]{Matthias~{Ivantsits}}
\author[17]{Jun~{Ma}}
\author[13]{Craig {Meyer}}
\author[18,19]{Rishabh {Sharma}}
\author[3]{Jixi {Shi}}
\author[19]{Nikolaos V. {Tsekos}}
\author[20]{Marta {Varela}}
\author[21]{Xiyue~{Wang}}
\author[22]{Sen~{Yang}}
\author[14]{Hannu~{Zhang}}
\author[23]{Yichi {Zhang}}
\author[24]{Yuncheng {Zhou}}
\author[24]{Xiahai {Zhuang}}
\author[3]{Raphael {Couturier}}
\author[1]{Fabrice {Meriaudeau}}

\cortext[cor1]{Corresponding author: Alain Lalande, alain.lalande@u-bourgogne.fr}
\fntext[fn1]{Alain Lalande and Zhihao Chen contributed equally to this work}
\address[1]{ImViA laboratory, University of Burgundy, Dijon, France}
\address[2]{MRI department, University  Hospital of Dijon, Dijon, France}
\address[3]{Femto-ST laboratory, University of Franche-Comté, Belfort, France}
\address[4]{Cardiology department, University  Hospital of Dijon, Dijon, France}
\address[5]{CASIS Company, Quetigny, France}
\address[6]{National Engineering School of Sousse, University of Sousse, Sousse, Tunisia}
\address[7]{LASEE laboratory, National Engineering School of Monastir, University of Monastir, Monastir, Tunisia}
\address[8]{Biomedical Imaging Group Rotterdam, Erasmus MC, Rotterdam, The Netherlands}
\address[9]{Department of Radiology and Nuclear Medicine, Erasmus MC, Rotterdam, The Netherlands}
\address[10]{Department of Computer Science, University of Copenhagen, Copenhagen, Denmark}
\address[11]{Centre of Marine Sciences, University of Algarve, Faro, Portugal}
\address[12]{School of Biomedical Engineering and Imaging Sciences, King’s College London, London, The United Kingdom}
\address[13]{Department of Biomedical Engineering, University of Virginia, Charlottesville, The United States}
\address[14]{Charit\'e – Universit\"atsmedizin Berlin, Berlin, Germany}
\address[15]{Fraunhofer MEVIS, Bremen, Germany}
\address[16]{German Centre for Cardiovascular Research, Berlin, Germany}
\address[17]{Department of Mathematics, Nanjing University of Science and Technology, Nanjing, China}
\address[18]{Data Analysis and Intelligent Systems Lab, Department of Computer Science, University of Houston, Houston, The United States}
\address[19]{Medical Robotics and Imaging Lab, Department of Computer Science, University of Houston, Houston, The United States}
\address[20]{National Heart and Lung Institute, Imperial College London, London, The United Kingdom}
\address[21]{College of Computer Science, Sichuan University, Chengdu, China}
\address[22]{College of Biomedical Engineering, Sichuan University, Chengdu, China}
\address[23]{School of Biological Science and Medical Engineering, Beihang University, Beijing, China}
\address[24]{School of Data Science, Fudan University, Shanghai, China}

}

\begin{abstract}
%%%
A key factor for assessing the state of the heart after myocardial infarction (MI) is to measure whether the myocardium segment is viable after reperfusion or revascularization therapy. Delayed enhancement-MRI or DE-MRI, which is performed several minutes after injection of the contrast agent, provides high contrast between viable and nonviable myocardium and is therefore a method of choice to evaluate the extent of MI. To automatically assess myocardial status, the results of the EMIDEC challenge that focused on this task are presented in this paper. The challenge's main objectives were twofold. First, to evaluate if deep learning methods can distinguish between normal and pathological cases. Second, to automatically calculate the extent of myocardial infarction. The publicly available database consists of 150 exams divided into 50 cases with normal MRI after injection of a contrast agent and 100 cases with myocardial infarction (and then with a hyperenhanced area on DE-MRI), whatever their inclusion in the cardiac emergency department. Along with MRI, clinical characteristics are also provided. The obtained results issued from several works show that the automatic classification of an exam is a reachable task (the best method providing an accuracy of 0.92), and the automatic segmentation of the myocardium is possible. However, the segmentation of the diseased area needs to be improved, mainly due to the small size of these areas and the lack of contrast with the surrounding structures. 
%%%%
\end{abstract}


\end{frontmatter}}]

%\linenumbers
\section{Introduction}

Myocardial infarction (MI) can be defined as myocardial cell death secondary to prolonged ischemia. %at the level of a segment of the myocardium. 
One crucial parameter to estimate the prognosis after myocardial injury and then to evaluate the state of the heart after MI is the viability of the considered segment, {\em i.e.} if the segment recovers its functionality upon revascularization. 
%Assessement of myocardial viability can be done from dobutamine stress echocardiography, from scintigraphy and MRI. 
From cardiac MRI, the viability can be evaluated thanks to the assessment of left ventricular end-diastolic wall thickness, the evaluation of contractile reserve, and the extent and the transmural nature of the infarction evaluated from delayed-enhancement MRI (DE-MRI) (\cite{schinkel2007, Kim1999}). %Selvanayagam et al. demonstrated that 
DE-MRI is a powerful predictor of myocardial viability after coronary artery surgery, suggesting an important role for this technique in clinical viability assessment ( \cite{selvanayagam2004}). 

A preliminary challenge organized in 2012 %by Karim et al. 
(\cite{Karim2016}) has been already dedicated to the automatic processing of DE-MRI. %Five groups took part in this challenge. 
This challenge showed promising results, but also indicated that some improvements should be done for potential use in clinical practice. The published dataset %(in 2012) 
was rather small (including fifteen human and fifteen porcine pathological cases), did not target a specific disease and no clinical data were associated. % Indeed, it was only composed of pathological cases, ,  

As part of the Emidec challenge (automatic Evaluation of Myocardial Infarction from Delayed-Enhancement Cardiac MRI, http://emidec.com/) organized during the MICCAI conference in 2020, the objective of our paper is first to compare the latest methodological developments in image processing, in particular deep learning approaches such as convolutional neural network (CNN), to automatically segment the DE-MRI exams (including normal and pathological cases with myocardial infarction and with or without persistent microvascular obstruction (PMO)), and secondly, thanks to the images and associated clinical data, to automatically classify the exams as normal or pathological. One of the main strengths of our database is the association of clinical data with DE-MRI, simulating the routine workflow in emergency departments.

%\section{Previous works}

\section{Materials and Methods}

\subsection{DE-MRI and clinical information}
The EMIDEC dataset contains patients admitted in cardiac emergency department with symptoms of a heart attack. This dataset is composed of 150~patients, each of which with a MRI exam and the associated clinical characteristics. The exam is a conventional one acquired at the University Hospital of Dijon (France) and done to study the left ventricle in case of symptoms of heart attack and is compound of kinetic and DE-MR images. For the DE-MRI, the images are acquired roughly 10 minutes after the injection of a gadolinium-based contrast agent. A series of short axis slices cover the left ventricle from the base to the apex, allowing the accurate evaluation of the extent of myocardial infarction. The pixel spacing is between $1.25\times 1.25$ mm\textsuperscript{2} and $2\times 2$ mm\textsuperscript{2} (according to the patient), with a  slice thickness of 8 mm and an image gap of 10 mm.
%/a distance between slices of 10 mm (i.e., o). 
The shift among slices due to the patient's breath hold was corrected and all the slices for one exam had been aligned according to the gravity centre of the epicardium. Along with the MRI, clinical and patient information were also recorded. Acquired data were fully anonymized and handled within the regulations set by the local ethical committee. As the data were collected retrospectively, and as the data are completely untraceable, for the French law, and for the staff of the ethical committee of the University Hospital of Dijon, it was not necessary to undergo the process of applying for an ethical approval number. In particular, concerning the images, using the NIfTi format, all the administrative information included in the header was discarded. Moreover, the clinical information is not specific enough to retrieve a patient. The patient features are characterised in Table \ref{tab_features} (\cite{Lalande2020}). A patient was considered overweight when the body mass index (BMI) is higher than 25. The history of coronary artery disease is positive if there is a previous acute cardiac event. The study of the electrocardiogram (ECG) allows classifying the heart attack as STEMI (ST-elevation myocardial infarction) type or not. STEMI-like myocardial infarction is the most serious type of heart attack, which is characterized by a long interruption of blood supply. A troponin test measures the levels of troponin T or troponin I proteins in the blood. These proteins are released during a myocardial infarction. Another biological marker is the NT-pro-brain natriuretic peptide (NT-proBNP) measured in venous blood, and it is an indicator for the diagnosis of heart failure (\cite{cochet2004}). The left ventricular ejection fraction (LVEF) is calculated in the emergency department from echocardiography during the reception of the patient. Finally, The Killip max corresponds to the maximum Killip score, which is a classification based on the physical examination of patients with possible acute MI (\cite{Killip1967}).

\begin{table*}[htbp]
\caption{Characteristics of the patient features for pathological and non-pathological cases.  %\The pathology diagnosis is according to the DE-MRI. 
}

\begin{center}

\begin{adjustbox}{width=0.7\textwidth,center}

\begin{threeparttable}[t]
\begin{tabular}{|c|c|c|}
\hline
Patient feature & Non-pathological subjects(n=50) & Pathological subjects(n=100) \\ \hline
Sex&38 females and 12 males &23 females and 77 males \\
Age& $66\pm14$ years&$59\pm12$ years\\
Tobacco (yes, no, former smoker)&18\%, 22\%, 60\%&44\%, 21\%, 35\%\\
Overweight\tnote{1}&62\%&53\%\\
Arterial hypertension&58\%&31\%\\
Diabetes&20\%&10\%\\
History of coronary artery disease& 4\%& 12\%\\
ECG (ST elevation)  &30\%& 80\%\\
Troponin (ng per mL)&$7.68\pm12.91$&$101.04\pm101.35$\\
Killip max (1,2,3,4)&76\%, 22\%, 2\%, 0\%&83\%, 12\%, 2\%, 3\%\\
LVEF\tnote{2}~ (percentage)&$49.62\pm13.49\%$&$47.74\pm13.17\%$\\
NTProBNP\tnote{3}~ (pg per mL)&$2136\pm3696$&$1314\pm2109$\\
\hline
\end{tabular}
\begin{tablenotes}
\item[1] If BMI %(Body Mass Index) 
  $>25$.
\item[2] Left Ventricular Ejection Fraction, calculated from transthoracic echocardiography.
\item[3] N-terminal pro-B-type natriuretic peptide.
\end{tablenotes}
\end{threeparttable}
\label{tab_features}
\end{adjustbox}
\end{center}
\end{table*}

\subsection{Dataset and contests}
The overall dataset consists of 150 exams, with 100 cases for the training (67~pathological cases and 33~normal cases, where ground truths are provided) and 50 cases for the testing (33~pathological cases, 17~normal cases). Each exam is divided into two parts, a DE-MRI exam composed of a series of short-axis slices and the associated clinical information (\cite{Lalande2020}). For each image, the contours of the myocardium, as well as the contours of the infarcted area and the PMO areas, if present, are considered as the ground truths, allowing the calculation of the main clinical metrics considering the whole slices for one exam. Tissue characteristics according to the manual annotations can be found in Table \ref{tab_MRI}. Along with MRI, the clinical and physiological characteristics are provided. The EMIDEC challenge contains two independent contests: the segmentation challenge and the classification challenge. The goal of the segmentation contest is to compare the performance of automatic methods on the segmentation of the myocardium for all the DE-MRI exams, as well as the segmentation of the myocardial infarction and PMO areas on exams classified as pathological ones. The goal of the classification contest is to classify the exams as normal or pathological, according to the clinical data with or without the DE-MRI exams (two sub-challenges).In order to avoid any bias between the two contests, the order of the cases is different in the testing set, and moreover, new cases were randomly added (and similarly some were removed) for the classification contest. 

\begin{table*}[htbp]
\caption{DE-MRI evaluations of pathological and non-pathological cases according to the manual annotations. \\
This table lists the characteristics of different tissues in the DE-MRI for the whole dataset.}

\begin{center}

\begin{adjustbox}{width=0.9\textwidth,center}
\begin{threeparttable}[t]

\begin{tabular}{|c|ccc|ccc|}
\hline
\multirow{2}{*} {Tissue} & \multicolumn{3}{c|}{Non-pathological subjects (n=50)}& \multicolumn{3}{c|}{Pathological subjects (n=100)}\\ \cline{2-7}
& Volume (cm\textsuperscript{3})\tnote{1}& PIM (\%)\tnote{2}& Presence (\%)\tnote{3}& Volume (cm\textsuperscript{3}) & PIM (\%) & Presence (\%)\\ \hline
Myocardium&96.32$\pm$22.07&-&-&119.28$\pm$32.28 &- &-\\
Left ventricular cavity&83.32$\pm$25.27&-&-&128.87$\pm$48.17&-&-\\
Myocardial infarction\tnote{4}&0&0&0&23.55$\pm$19.28&18.25$\pm$11.52&100 (79.78)\\
PMO&0&0&0&2.34$\pm$5.14&1.65$\pm$3.03&51 (23.27)\\
\hline
\end{tabular}
\begin{tablenotes}
\item[1] Absolute tissue volume per case.
\item[2] Percentage of Infarcted Myocardium. This index is reserved for myocardial infarction and PMO.
\item[3] Percentage of cases where tissue is present, while the value in brackets gives the percentage of slices.
\item[4] The PMO is included.
\end{tablenotes}

\end{threeparttable}
\label{tab_MRI}
\end{adjustbox}
\end{center}
\end{table*}

\subsection{Evaluation metrics}
For the segmentation contest, the clinical metrics are the most widely used in cardiac clinical practice, {\em i.e.} the average errors for the volume of the myocardium of the left ventricle (in cm\textsuperscript{3}), the volume (in cm\textsuperscript{3}) and the percentages of MI and PMO. The geometrical metrics are the average Dice index for the different areas and the Hausdorff distance (in 3D) for the myocardium.  For each metric, a ranking is done, and the final ranking consists of the sum of the ranking for each metric. To better evaluate the segmentation results of the PMO, the case-wise and slice-wise accuracies are additionally calculated, but were not taken into account for the challenge ranking. For the classification contest, only classification accuracy was used.

%For the segmentation contest, the clinical metrics are those that are the most widely used in cardiac clinical practice, {\em i.e.} the average errors for the volume of the myocardium of the left ventricle (in mm\textsuperscript{3}), and the average error of the volume (in mm\textsuperscript{3}) and the percentage of MI and of PMO. The geometrical...

\section{Evaluated architectures}

\subsection{Segmentation contest}
The main objective of the segmentation contest is to automatically provide the contours of the myocardium on each slice, as well as the delineation of the diseased areas. % (if present).

\subsubsection{Image preprocessing and data augmentation}
To ensure that semantic information in DE-MRI can be efficiently interpreted by the segmentation models, some challengers employed adaptive image preprocessings on the raw MR images. For example, image normalization aims at correcting the heterogeneous intensity between cases. \cite{Yang} and \cite{Feng} applied the Z-score normalization on each slice with the following formula: 
\begin{equation}
    z=\frac{x-\mu}{\sigma}
\end{equation}
where $z$ is the pixel intensity after the Z-score normalization, $\mu$ the mean intensity at the level of the MR slice and $\sigma$ the standard deviation of the slice intensity. Normalized images have a grey level distribution with zero mean and unit standard deviation so that the inter-case intensity distribution is uniform.

Exams have some slight inconsistencies in the plane dimensions. In order to ensure a uniform input shape of the predictive models, challengers processed the plane dimensions of the input data differently. The first type of method is cropping, {\em e.g.} \cite{Feng} cropped a fixed size in the centre of each slice. In addition to the cropping, a linear interpolation was also performed to resize the images to a uniform shape (\cite{Camarasa}). Besides the processing on the slice shape, some challengers also interpolated the image to have a consistent pixel spacing.  Thanks to the alignment of the slices according to the gravity center of the epicardium, no additional preprocessing concerning the relative inter-slice position is needed if a 3D predictive model is employed by the challengers.

The amount of training data directly affects the performance of supervised models. A reasonable data augmentation method can equivalently expand the size of the training set. \cite{Camarasa} performed rotations, elastic deformations, and flips on slices to randomly produce supplementary training data while \cite{Feng} forced the model to ignore the specificity for different orientation features by the rotations only. \cite{Lourenco} adjusted the original semantic information by adding stochastic noise, applying k-space corruption, small image rotations, intensity scalings, and smooth non-rigid deformations. \cite{zhou} proposed another data augmentation method that was based on the mix-up strategy (\cite{mixup}). The mix-up strategy constructs virtual training examples as follows:
\begin{gather}
\tilde x = \lambda x_{i}+(1-\lambda)x_{j}\\
\tilde y = \lambda y_{i}+(1-\lambda)y_{j}
\end{gather}
where $x_i$ and $x_j$ are raw input vectors, $y_i$ and $y_j$ are one-hot label encodings, $\tilde x$  and $\tilde y$ is the pair of artificially  created data. $\lambda$ is a coefficient belonging to $[0,1]$. Based on this approach, Zhou et al. made a targeted improvement to make the generated images closer to a blend of two adjacent images. The proposed mix-up formula for the MRI augmentation is:
\begin{equation}
\tilde x =\lambda x_i+(1-\lambda)Tx_j
\end{equation}
where $T$ denotes an affine transformation, and accordingly the similar formula for the mask data augmentation. Given the greater focus on the ROI (Region Of Interest corresponding to the myocardium), the affine transformation $T$ tries to fit the transformation from the foreground area (LV+Myocardium) in a randomly chosen slice $x_i$ to the foreground area in another randomly chosen slice $x_j$. In the affine transformation, the scaling factor, {\em i.e.} the linear map is $[s,s]^\top$ where $s=l^i / l^j$, $l^i$ and $l^j$ are the average distance from the foreground pixels to the foreground center for the slice $i$ and the slice $j$, respectively. The translation offset is $[c_x^i - c_x^j,  c_y^i - c_y^j]^\top$ where $c_x$ and $c_y$ denote the coordinates of the foreground area centre. Thus, the matrix of $T$ is:
\begin{equation}
\begin{pmatrix}
 s&  0& c_x^i - s\cdot c_x^j\\ 
 0&  s& c_y^i - s\cdot c_y^j\\ 
 0&  0& 1
\end{pmatrix} 
\end{equation}
\subsubsection{Segmentation frameworks}
\begin{figure*}[htbp]
\includegraphics[width=1\textwidth]{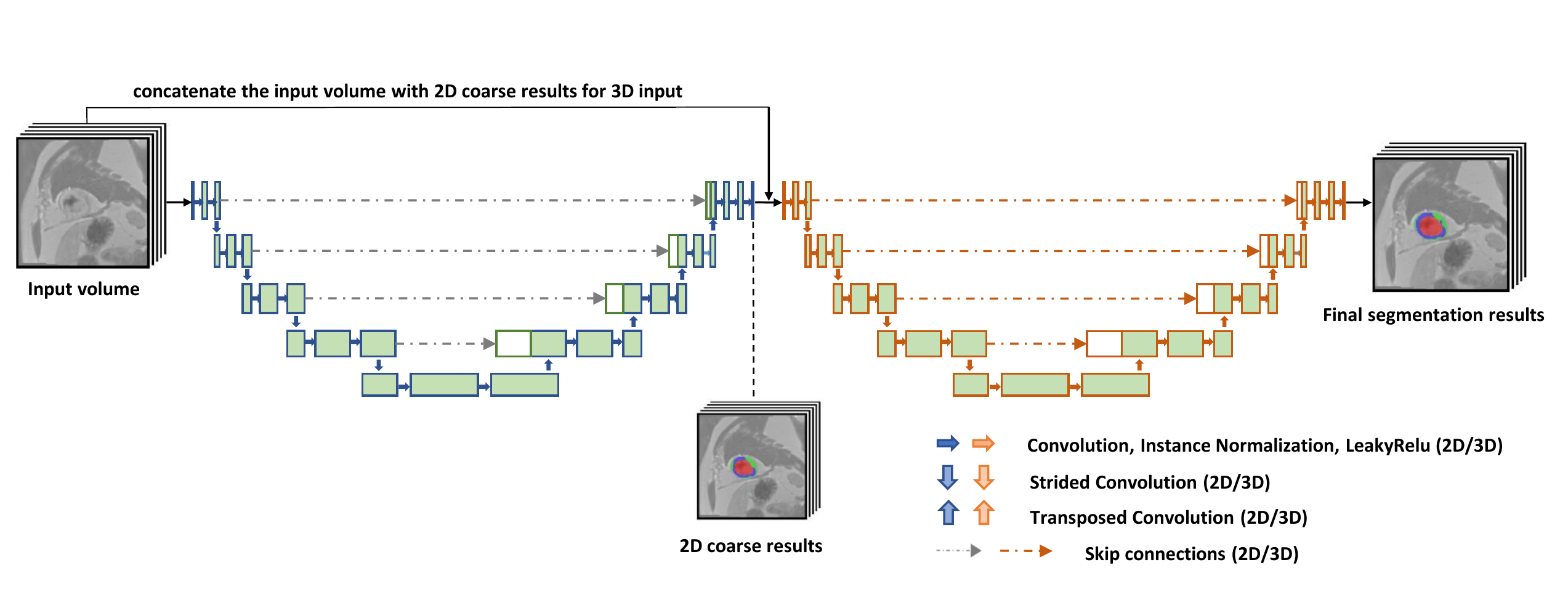}
\caption{Cascaded 2D-3D framework of \cite{Zhang} for the myocardial tissue segmentation. The 2D network on the left performs a preliminary  segmentation focusing on intra-slice information. The 3D network on the right then takes the MRI and the obtained segmentation mask as its input. 
}
\label{network_zhang}
\end{figure*}

%Challengers employed segmentation frameworks of different stages. Most of participants first delineated the myocardium (endocardial and epicardial borders), and then the different myocardial tissues were segmented in the ROI corresponding to the myocardium with another model in a second step, while other participants proposed one-stage models to obtain the end-to-end segmentation of all the target tissues.
Challengers employed segmentation frameworks with a different number of stages. Most of the challengers first delineated the myocardium (endocardial and epicardial borders) and then segmented the different myocardial tissues in the ROI corresponding to the myocardium with another model in a second step. Other challengers proposed one-stage models to obtain an end-to-end segmentation of all the target tissues.
%It is interesting to point some highlights of the framework conceptions. 
\cite{Zhang} proposed the cascaded 2D-3D framework where the 2D-model's receptive field was limited to intra-slice for preliminary segmentation, then the cascaded 3D-model took the 2D preliminary segmentation mask and the whole volume for the fine segmentation. This conception aims at restricting the impact of intra-slice heterogeneity and taking into account the volumetric information for the more accurate segmentation. The networks' configurations are inspired by nnUnet (\cite{nnUnet}). Figure~\ref{network_zhang} shows the architecture of the cascaded two-stage framework. \cite{Camarasa} also employed a usual two-stage segmentation pipeline but the scar segmentation was uncertainty-based: the ROI segmented by the first model passed through a probabilistic Auto-Encoder using Monte-Carlo dropout. The generated uncertainty map corresponding to the segmented ROI by the Auto-Encoder was then fed into the second model for the scar segmentation. This proposal was intended to assess whether the method could increase the attention on rare examples that are otherwise poorly segmented.

\subsubsection{U-Net-based encoding-decoding models}
The semantic image segmentation task can be usually treated with encoding-decoding models. Most challengers employed U-Net-based models (\cite{Unet}) motivated by its success in many medical image segmentation work. In this subsection, the details of all the employed U-Net-based models will be introduced.
\paragraph{Building blocks}
\label{buildingblocks}
The vanilla U-Net employs the conventional convolution-pooling architecture as the basic encoding block. To better interpret the semantic information, challengers attempted with more recent blocks of CNNs. In the encoding branch, \cite{Yang} and \cite{Girum} applied the Squeeze-and-Excitation (SE) block (\cite{SE}) to better model the interdependencies between channels of the convolutional features. To this end, in the SE block, feature maps were first squeezed into a channel descriptor with shape [1, 1, channel] by the global average pooling. Then to fully capture the aggregated channel-wise information, a simple gating mechanism was employed with linear transformations and non-linear activation functions:
\begin{equation}
    \mathbf{F}_{ex}(\mathbf{z, W}) =  \rho(\mathbf{W}_2\delta(\mathbf{W}_1\mathbf{z}))
\end{equation}
where $\mathbf{W}_1\in \mathbb{R}^{\frac{C}{r}\times C}$ and $\mathbf{W}_2\in \mathbb{R}^{C\times \frac{C}{r} }$ are linear transformations, $C$ and $r$ are channel size and reduction rate, $\delta$ refers to ReLU (\cite{ReLU}) and $\rho$ refers to sigmoid activation. To finally emphasize differently the feature maps, the channel-wise multiplication was operated between the scaled squeeze-excitation scalar and the feature maps. The SE block can be combined with other convolution architectures since it aims at providing additional interdependencies between the features maps obtained from convolution blocks.

On the decoding side, Inverted Residual Blocks (IRB) were employed by \cite{Brahim}. The IRB has been proposed by \cite{MobileNet} in MobileNetV2. It consists in a series of $1\times 1$ convolution, depth-wise $3\times 3$ convolution and $1\times 1$ convolution, and the skip connection. The IRB follows an inverse order of the feature map number comparing to the original Residual block (\cite{resnet}). In IRB the network is expanded by the first  $1\times 1$ convolution and squeezed by the second $1\times 1$ convolution. This conception was initially intended for lightweight network for mobile applications thanks to the reduced number of parameters of the depth-wise convolution.

Selective Kernel (SK) (\cite{SKnetwork}) was another block employed in the decoding side by \cite{Yang}. The SK block aims to adaptively adjust the receptive field sizes. To enable the automatic kernel size selection, three operators are used in SK: split, fuse, and select. The split operator creates two branches for the next operators where the first branch passes through a conventional $3\times 3$ convolution and another is a $3\times 3$ dilated convolution with a dilation size of~2. Then, in the fuse operator, a third branch is created to store the multi-kernel information. In this branch, the feature maps obtained by the split operator are first fused by element-wise summation, and then embedded by global average pooling. A fully-connected layer compacts the fused feature into a lower dimension. Finally, in the select operator, the compact feature guides the selection of different spatial scales of information for the feature maps of the first and second branches by soft attention across channels. The definitive output of the SK block is the sum of the first and second branches considering the attention weights achieved by the soft attention across channels.

\begin{figure*}[htbp]
\includegraphics[width=1\textwidth]{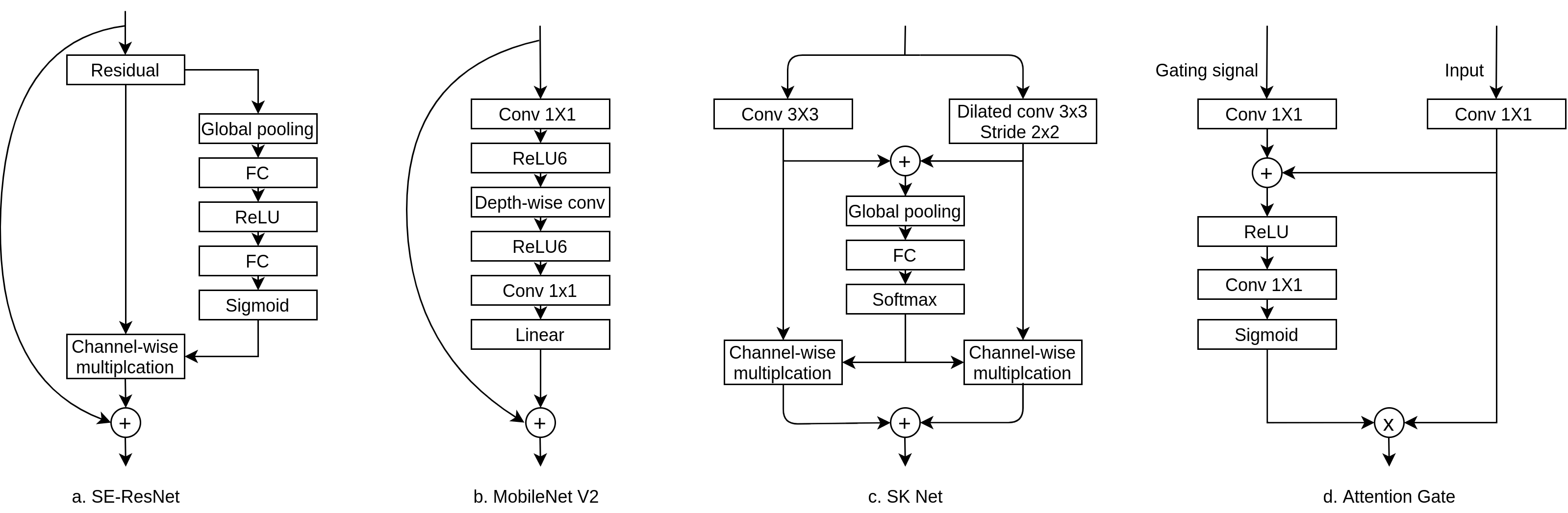}
\caption{Featured building blocks for the segmentation networks. The conceptions come from the original papers of the challengers or slightly modified to better adapt the contest. a. SE-ResNet: the residual SE block, b. MobileNet V2: the IRB from MobileNet V2, c. SK Net: the SK module can be deployed in the encoding or decoding phases, d. Attention Gate: the attention gating from Attention U-Net should be deployed at the skip connection. The Gating signal comes from the encoding side and the input signal denotes the up-sampled features from the decoding side. The first two conv 1x1 ensure the same number of channels for the two signals of the Attention Gate.
}

\label{blocks}
\end{figure*}

In addition to the above-featured building blocks, challengers also tried other relatively more common blocks such as the residual block of the ResNet (\cite{resnet}) and its aggregated variant ResNeXt (\cite{ResNeXt}), as well as the Inception module where convolutions of different receptive field interpret input features at the same time (\cite{inception}). The attention block (\cite{attentionunet}) was also mentioned by several challengers to focus on valuable features at the skip connections and the bottleneck. The illustration of the featured building blocks can be found in Figure \ref{blocks} .

\begin{figure}[htbp]
\centering
\includegraphics[width=0.5\textwidth]{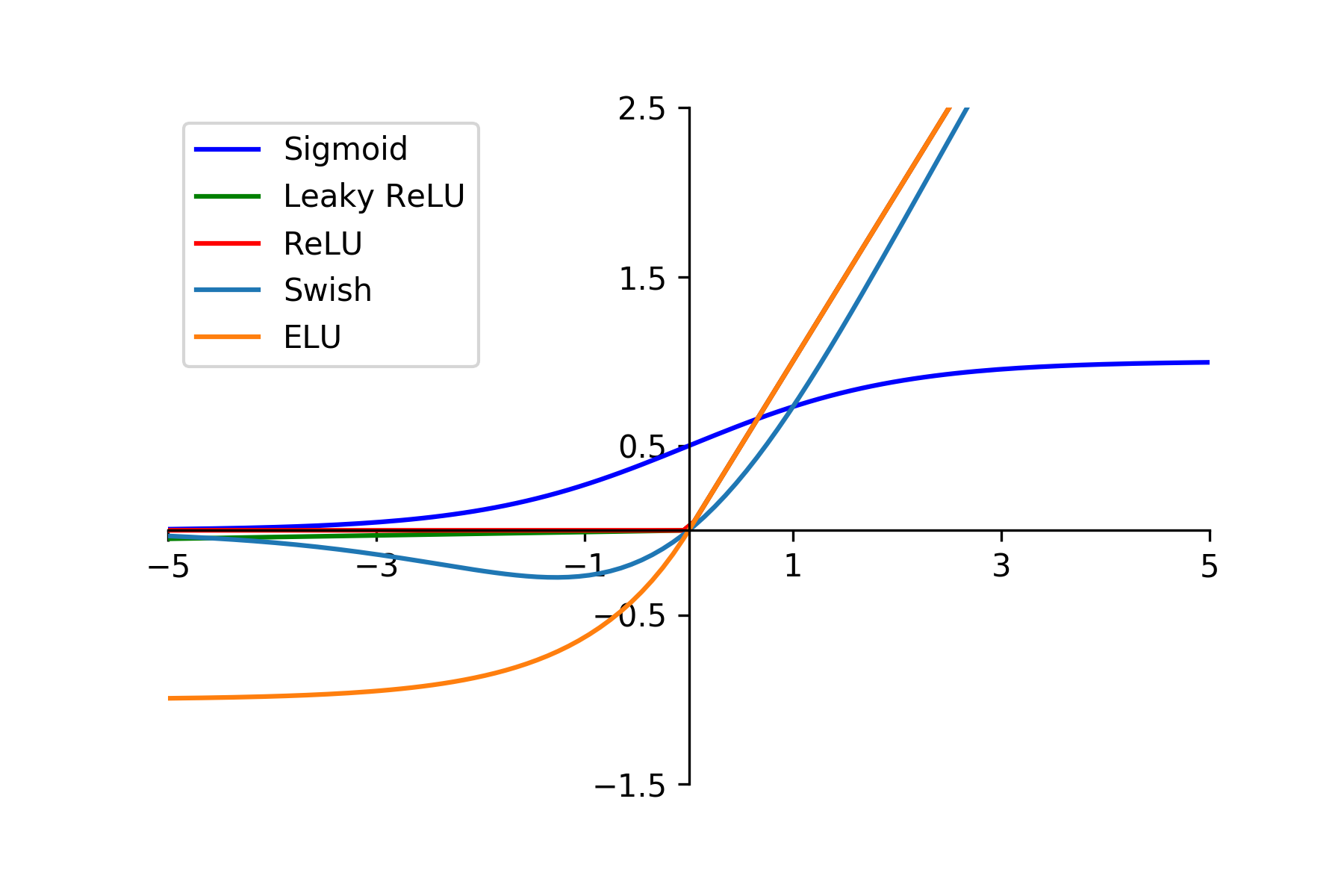}
\caption{Deployed non-linear activation functions in the neural networks for the segmentation task.
}
\label{activation}
\end{figure}

Challengers also reported the use of a variety of activation functions. Like in most of the current deep learning models, the activation functions themselves are all nonlinear equations, their core functionality is to ensure that nontrivial problems can be fitted by deep neural networks. Therefore, sigmoid, rectifier (ReLU) (\cite{ReLU}) and its leaky variant, exponential linear (ELU) (\cite{ELU}),  Swish (\cite{Swish}), {\em etc.} activation functions were employed. Figure \ref{activation} illustrates the deployed activation functions by challengers.

\paragraph{Loss functions and penalizations}
The category imbalance is significant in the challenge dataset, that is, the myocardial infarction and the PMO have few instances in terms of the number of pixels as shown in Figure \ref{tab_MRI}. To address this issue, challengers investigated different loss functions. Cross entropy loss calculates the average of the number of bits needed between the target and the prediction masks to identify the event of the automatic segmentation. The original cross entropy loss considers equally the instances of each class. A feature that in our case study causes the category imbalance in the prediction: the neural network will tend to predict all the pixels as the background class since it is easier to describe between the label and prediction distributions with a few necessary digits. To overcome the category imbalance, the weighted cross entropy evaluates differently the prediction pixels: the pixels of minor classes will have a more important contribution to the loss. The weights for each class can be set by calculating the inverse ratio of the number of instances under each class in the training set. The categorical weighted cross entropy loss can be formulated as:
\begin{equation} 
\label{eq_ce}
-\frac{1}{N} \sum_{l=1}^L \displaystyle w_l \sum_n r_{ln}log(p_{ln})
\end{equation}
where $r$ and $p$ denote the target and the prediction of the pixel $n$, $l$ denotes the class, the class weight is ${\displaystyle w_{l}}=1/\left (\sum_{n=1}^{N}r_{ln}  \right )$ where $N$ denotes all the images of the training set. Dice loss is another usual loss function for the segmentation task. It calculates the overlap between the target and the prediction comparing their surface. For binary segmentation, the Dice loss avoids the parameterization of weights since the randomness of the class appearance has been taken into account. However, for the multi-class task, the weights should be also calculated. The formula of the weighted multi-class Dice loss, termed as Generalized Dice (\cite{gdice}) is:
\begin{equation} 
\label{eq_Dice}
1-2\frac{\sum_{l=1}^{L} {\displaystyle w_{l}}\sum _{n} r_{ln}p_{ln}}  
{\sum_{l=1}^{L} {\displaystyle w_{l}}\sum _{n} r_{ln}+p_{ln}}
\end{equation}
where the weight $\displaystyle w_l$ is ${\displaystyle w_{l}}=1/\left (\sum_{n=1}^{N}r_{ln}  \right )^2$. %In fact, 
It can be observed that for the scar segmentation, the categorical cross entropy loss %employed by challengers 
was weighted (\cite{Zhang,Yang,zhou}) while the multi-class Dice loss was not weighted \cite{Zhang,Yang,Camarasa}. To leverage the cross entropy loss and Dice loss, their combination termed Comboloss (\cite{ComboLoss}) was also practiced by many challengers (\cite{Zhang,Yang,Girum}) for the ROI or the whole tissues segmentation. 

Apart from the loss functions that penalize the difference between the target and the prediction, other prior information-based penalizations were investigated by challengers. \cite{Brahim} applied the 3D auto-encoder as a part of the loss to refine the mask contours. The employment of the auto-encoder with the cardiac MRI was first proposed by \cite{autoencoder} for the myocardium segmentation. In the original work, the auto-encoder learns the 2D shape prior of the myocardium since the short-axis view of the left ventricle should be a closed circle except for the extreme apical and basal slices. The auto-encoder can be thought as an annex network following the segmentation network so that the loss of the auto-encoder takes part of the backpropagation. Similarly, with reference to the prior anatomical knowledge, \cite{zhou} proposed the neighborhood penalty as a weak constraint strategy. Given the fact that the PMO should be in contact with the infarction and the whole scar area should be inside the myocardium, this penalty encourages such correlated tissues to stick together.  
 
\paragraph{Inter-slice and intra-slice information}
The cardiac MR images can be considered as pseudo 3D data, {\em i.e.} the voxel spacing is inconsistent between the in-plane %(cross-section of the ventricle) 
and between planes. 
%Given the more coarse inter-slice correlation, c
Challengers adopted different strategies to deal with the inter- and intra-slice information. The first one omits the inter-slice correlation, {\em i.e.} all the tissues are segmented from single slices whether the framework is one-stage or two-stage (\cite{Huellebrand,Girum,zhou,Feng}). The second one only takes 3D inputs while the data format organization is different. \cite{Camarasa} employed a 3D CNN where the convolution kernel was 3D. \cite{Yang} treated multi-slice data as different channels, {\em i.e.} at the input layer each channel stocked one slice and the following convolutions were 2D. The major difference of the 3D data interpretation between the 3D convolution and the 2D convolution with RGB channels-like inputs is the relative positional information between the slices. The 2D convolution cannot distinct the slice order while the 3D convolution retains the inter-slice information as local vector data. The last strategy is a compromise approach (\cite{Zhang,Brahim}): the ROI or preliminary segmentation only refers to the intra-slice information and the scar or final segmentation considers both the intra- and inter-slice information. The purpose is to avoid the potential inter-slice heterogeneity for the myocardium or preliminary segmentation, and take the advantage of the inter-slice information for the scar since the recognition of different myocardial tissues relies more on their neighbouring slices.

\subsubsection{Mixture model for the scar segmentation}
Apart from the U-Net-based models that most challengers employed, a mixture model was proposed by \cite{Huellebrand} for the scar segmentation. The application of the mixture model on the cardiac MRI was inspired by the work of \cite{mixturemodel}. The mixture model differs the scar tissues only according to the intensity distribution. The challengers trialed the mixture of a Rician and a Gaussian distribution and the mixture of Rayleigh and Gaussian distribution, and then adopted the latter which was proved better fitted to the scar tissues in the DE-MRI. Finally, inspired by \cite{watershed}, a watershed segmentation in high-intensity voxels was used at the inner part of the myocardium to get the segmented contours.

\subsubsection{Post-processing}
According to prior information, challengers employed simple post-processing methods. \cite{Huellebrand} proposed a thresholding for the segmented PMO: assuming that the PMO should be in contact with the cavity or the infarction, the contours detected by morphological closing were removed from the raw segmentation of the PMO. \cite{Zhang} adopted another simple treatment that removed all the scattered pixels from the segmentation. 

\subsection{Classification contest}
The objective of the classification contest is to classify each exam as normal or pathological, whatever the extent of the myocardial infarction.
\subsubsection{Basic data interpretation algorithms}
Challengers employed a variety of machine learning-based algorithms to interpret the DE-MRI and the clinical features. Provided with the MRI, a simple down-sampling CNN as AlexNet (\cite{alexnet}) encodes the images to regression or classification outputs (\cite{Sharma,Shi, Lourenco}), or optionally U-Net based down-sampling up-sampling models yield the segmentation of different myocardial tissues so that the volume of each tissue can be quantified (\cite{Lourenco, Girum}). 

To interpret the textual data of the clinical and physiological information, the choice of predictive models is more diverse. The common functionality %of such models 
is its ability to solve non-linearly separable problems. For example, the MultiLayer Perceptron (MLP) (\cite{mlp}) is a feedforward artificial neural network. Inputs are passed through multiple layers in which data are mapped with non-linear activation functions in the forward stage (\cite{Ivantsits, Sharma}). The decision tree (\cite{decision_trees}) and the random forest (\cite{random_forest}) are flow-chart-like decision models that consist of nodes (\cite{Sharma, Shi,Ivantsits}). %In the decision tree, each node examines for an attribute and each branch represents the output of a test. Therefore, the branch where all the nodes are satisfied points to the prediction of the input. As its name suggests, the random forest is an optimisation of the decision tree consisting of a stack of trees. 
The random forest corrects the overfitting habit of the decision trees by training uncorrelated trees and the final decision is made by individual trees. Boosting methods are the ensemble of sequentially connected weak learners (\cite{boosting}). In the context of decision trees, the gradient boosting decision trees build a series of trees, which are the weak learners in this boosting method. Errors are passed between trees, with each tree attempting to reduce the errors passed from the previous tree (\cite{gradient_boosting_trees}) (\cite{Ivantsits}). Moreover, usual statistical models such as Support Vector Machine with non-linear kernel (\cite{kernel_tricks}), k-Nearest Neighbors (\cite{KNN}), the logistic regression (\cite{logisticregression}) were investigated by challengers (\cite{Sharma, Girum, Ivantsits}).

\subsubsection{Data fusion and decision about the presence of myocardial infarction}
\begin{figure*}[htbp]
%\hspace{1cm}
\includegraphics[width=0.95\textwidth]{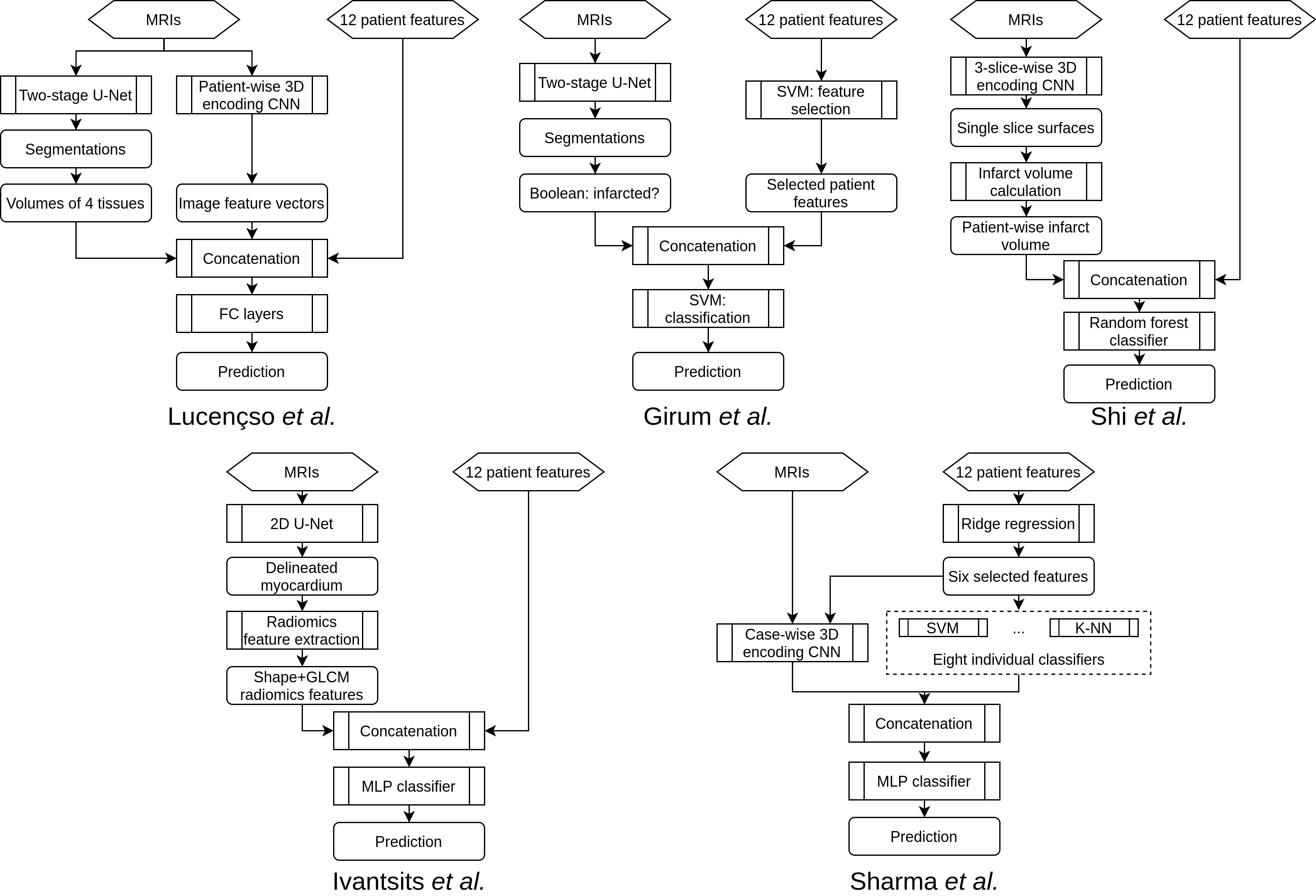}
\caption{Proposed multi-input classification pipelines by challengers for the classification contest.
}

\label{classification_fig}
\end{figure*}

The classification contest allows challengers to take  advantages of both the DE-MRI and the clinical and physiological data to make the automatic decision. However, the different format and dimension between the images and the textual data constrain the decision with a single predictive model. Data fusion is therefore a challenging issue to achieve the maximum semantic information. \cite{Lourenco}, \cite{Girum} and \cite{Shi} deployed the same strategy of predicting the volumes of different tissues as additional textual features alongside the 12 clinical and physiological features. Nevertheless, the volume estimation and the decision making models are different among these approaches. Lourenço et al. and Girum  et al. employed U-Net-based models to get the segmentation, while Shi  et al. performed an encoding CNN to directly get the surface regression. Apart from the surface regression methods, the concatenation of the surface information to other textual features was also variable. Lourenço  et al. added the volumes of all myocardial tissues as four additional textual features. Girum  et al. only considered if the case is pathological as one additional Boolean feature and Shi  et al. referred to the infarction volume as one additional numerical feature. \cite{Ivantsits}  tried to interpret the DE-MRI as textual information as well. However, the obtained textual information was radiomic features (\cite{radiomics}) instead of the volume of the tissues. The radiomic features interpreted from the DE-MRI were intended to model the myocardial features such as the intensity, shape, and spatial characteristics. In practice, Ivantsits et al. investigated the shape and the Gray Level Co-occurrence Matrix (GLCM) that described the second-order joint probability function of an image region as the experimental radiomic features. \cite{Sharma} proposed a stacked multi-modal approach without obtaining intermediate data such as the infarct volume or the radiomics features. The classifications were first achieved by a series of statistical models and a multi-modal CNN. Then the individual classifications were fed into an MLP to get the final decision. The application of the series of classification models could be thought as a boosting method and the models inside played the role of weak classifiers since their decisions would be judged together with the CNN's output by the MLP at the end of the proposal. All the diagrams of the classification pipelines can be found in Figure \ref{classification_fig}.

\section{Results}
%Although participants have reported their experimental results on the training dataset, in this section the results will refer to the segmentations and classifications the organizers received during the final challenge session. 
The results were obtained on the datasets used during the testing phase of the final challenge session.

\subsection{Segmentation contest}
%For quick reference, 
Table~\ref{resume_segmentation} shows the key conception details of the segmentation contest challengers. In Table \ref{res_segmentation} the evaluation results of each target tissue are provided. Results reveal that the myocardium segmentation is globally satisfying while the infarction is relatively challenging to be correctly predicted. The metrics of Dice and volumes used during the challenge for the PMO segmentation may not be consistent since the PMO only represents a very small volume of the data. Indeed, a total absence of PMO on all the images seemingly provides correct results with Dice index or volumes. In contrast, the accuracy highlights the efficiency of the different methods to detect PMO areas. Moreover, segmentation results by slice position can be found in the supplementary material.

\begin{table*}[htbp!]
\caption{Principal concepts of the methods for the segmentation contest.}

\begin{center}

\begin{adjustbox}{width=\textwidth,center}
\begin{threeparttable}[t]

\begin{tabular}{cccc}
\hline
%\begin{tabular}{@{}c@{}} \\ \end{tabular}

Challenger(s)& Framework & Methods & Highlights\\ \hline
Brahim  et al.&Two-stages&\begin{tabular}{@{}c@{}}Myocardium: 2D U-Net with Attention and IRB \\ Infarct: 3D U-Net variant\end{tabular}&3D Auto-encoder to perfect myocardium shape\\
Camarasa  et al.& Two-stages&\begin{tabular}{@{}c@{}}Myocardium: 3D U-Net variant \\ Infarct: 3D U-Net variant\end{tabular}  & Uncertainty myocardial area generated by probabilistic auto-encoder for infarct segmentation\\
Feng  et al.&One-stage & 2D U-Net with dilated convolution & Data augmentation with additional scar tissues\\
Girum  et al.& Two-stages&\begin{tabular}{@{}c@{}}Myocardium: 2D U-Net with SE block \\Infarct: 2D U-Net with SE block \end{tabular}&Independent myocardium and infarct segmentation from non-cropped MRI\\
Huellebrand  et al.& Two-stages&\begin{tabular}{@{}c@{}}Myocardium: 2D U-Net variant \\Infarct: mixture model \end{tabular}&\begin{tabular}{@{}c@{}}Transfer learning with cine-MRI \\Post-processing with thresholding and morphological closing \end{tabular} \\
Yang  et al.& One-stage& 2D U-Net with SE and SK blocks&\begin{tabular}{@{}c@{}} RGB channel-like adjacent slices input\\ Two decoder branches supervised by myocardium and infarct masks\end{tabular} \\
Zhang &  Two-stages&\begin{tabular}{@{}c@{}}Preliminary: 2D U-Net variant \\ Definitive: 3D U-Net variant\end{tabular}  &3D MRI with cascaded 2D segmentation as 3D input \\
Zhou  et al.&One-stage& 2D U-Net with Attention& \begin{tabular}{@{}c@{}}Data augmentation with mix-up strategy \\Neighborhood penalty as neighboring loss \\\end{tabular}\\

\hline
\end{tabular}

\end{threeparttable}
\label{resume_segmentation}
\end{adjustbox}
\end{center}
\end{table*}

\begin{table*}[htbp!]
\begin{center}
\caption{Results of the segmentation contest. The metrics are given by target tissue (myocardium, infarct and PMO). The table is sorted by the general ranking of the contest, which is calculated from the nine subranks. Best results in bold.}
\begin{adjustbox}{width=1\textwidth}
\begin{threeparttable}[t]
\begin{tabular}{c|*{3}{c}|*{3}{c}|*{5}{c}}
\hline
\multirow{2}{*}{\textbf{Challenger(s)}}& \multicolumn{3}{c|}{\textbf{Myocardium}}&\multicolumn{3}{c|}{\textbf{Infarction}}&\multicolumn{5}{c}{\textbf{PMO}}\\ \cline{2-12}
&Dice&Vol. Diff. (cm$^3$)& Hausdorff (mm)& Dice& Vol. Diff. (cm$^3$)& Pct. Diff. (\%)\tnote{1}&Dice& Vol. Diff. (cm$^3$)& Pct. Diff. (\%)\tnote{1}& Acc. (case,\%)\tnote{2}&Acc. (slice,\%)\tnote{2}\\ \hline
Zhang & \textbf{0.879}$\pm$0.027& \textbf{9.26}$\pm$9.08 & \textbf{13.01}$\pm$8.81 & \textbf{0.712}$\pm$0.268& \textbf{3.12}$\pm$5.15 & \textbf{2.38}$\pm$0.031 & \textbf{0.785}$\pm$0.393& \textbf{0.63}$\pm$2.27 & \textbf{0.38}$\pm$0.012 & \textbf{84.00}&\textbf{94.97}    \\
Feng  et al. & 0.836$\pm$0.124& 15.19$\pm$16.41 & 33.77$\pm$111.63& 0.547$\pm$0.340& 3.97$\pm$8.36 & 2.89$\pm$0.045 & 0.722$\pm$0.432& 0.88$\pm$3.41 & 0.53$\pm$0.017 &80.00& 90.78    \\
Yang  et al. & 0.855$\pm$0.027& 16.54$\pm$10.27 & 13.23$\pm$6.80 & 0.628$\pm$0.315& 5.34$\pm$7.88 & 4.37$\pm$0.062 & 0.610$\pm$0.463& 1.85$\pm$3.32 & 1.69$\pm$0.033 &76.00 & 81.56   \\
Huellebrand  et al. & 0.841$\pm$0.051& 10.87$\pm$8.53 & 18.3$\pm$15.74 &0.379$\pm$0.296& 6.17$\pm$8.36 & 4.93$\pm$0.059& 0.523$\pm$0.483& 0.95$\pm$3.00 & 0.64$\pm$0.015 &70.00&85.75\\
Camarasa  et al. & 0.757$\pm$0.111& 17.11$\pm$15.45 & 25.44$\pm$21.71 & 0.308$\pm$0.280& 4.87$\pm$8.49 & 3.64$\pm$0.047&0.605$\pm$0.485& 0.87$\pm$3.27 & 0.52$\pm$0.016&74.00&84.36  \\
Zhou  et al.&0.825$\pm$0.057& 13.29$\pm$11.34 & 83.42$\pm$158.97  & 0.378$\pm$0.309& 6.10$\pm$9.45 & 4.71$\pm$0.06 & 0.520$\pm$0.487& 0.88$\pm$3.38 &  0.54$\pm$0.017& 64.00&86.87   \\ 
Brahim  et al.\tnote{3}& 0.791$\pm$0.050& 12.68$\pm$10.59 &23.87$\pm$11.52& 0.274$\pm$0.379& 7.05$\pm$12.73 & 5.19$\pm$0.074 &0.641$\pm$0.479& 0.83$\pm$3.109 & 0.50$\pm$0.016 &74.00&89.39     \\
Girum  et al.\tnote{3} & 0.803$\pm$0.057& 11.81$\pm$14.09 & 51.48$\pm$98.15&  0.340$\pm$0.474& 11.52$\pm$16.53 & 8.58$\pm$0.101 &0.780$\pm$0.414& 0.89$\pm$3.61 & 0.51$\pm$0.018 &78.00&89.66   \\\hline
\end{tabular}
\begin{tablenotes}
\item[1] Pct. Diff. : Difference between the percentage of the infarcted myocardium.
\item[2] Additional metrics. These metrics were not taken into account in the ranking.
\item[3] Co-author(s) come(s) from the challenge organization team. Do(es) not participate in rankings.
\end{tablenotes}
\end{threeparttable}
\label{res_segmentation}
\end{adjustbox}
\end{center}
\end{table*}

\begin{figure*}[htbp!]
\hspace{1cm}
\includegraphics[width=0.8\textwidth]{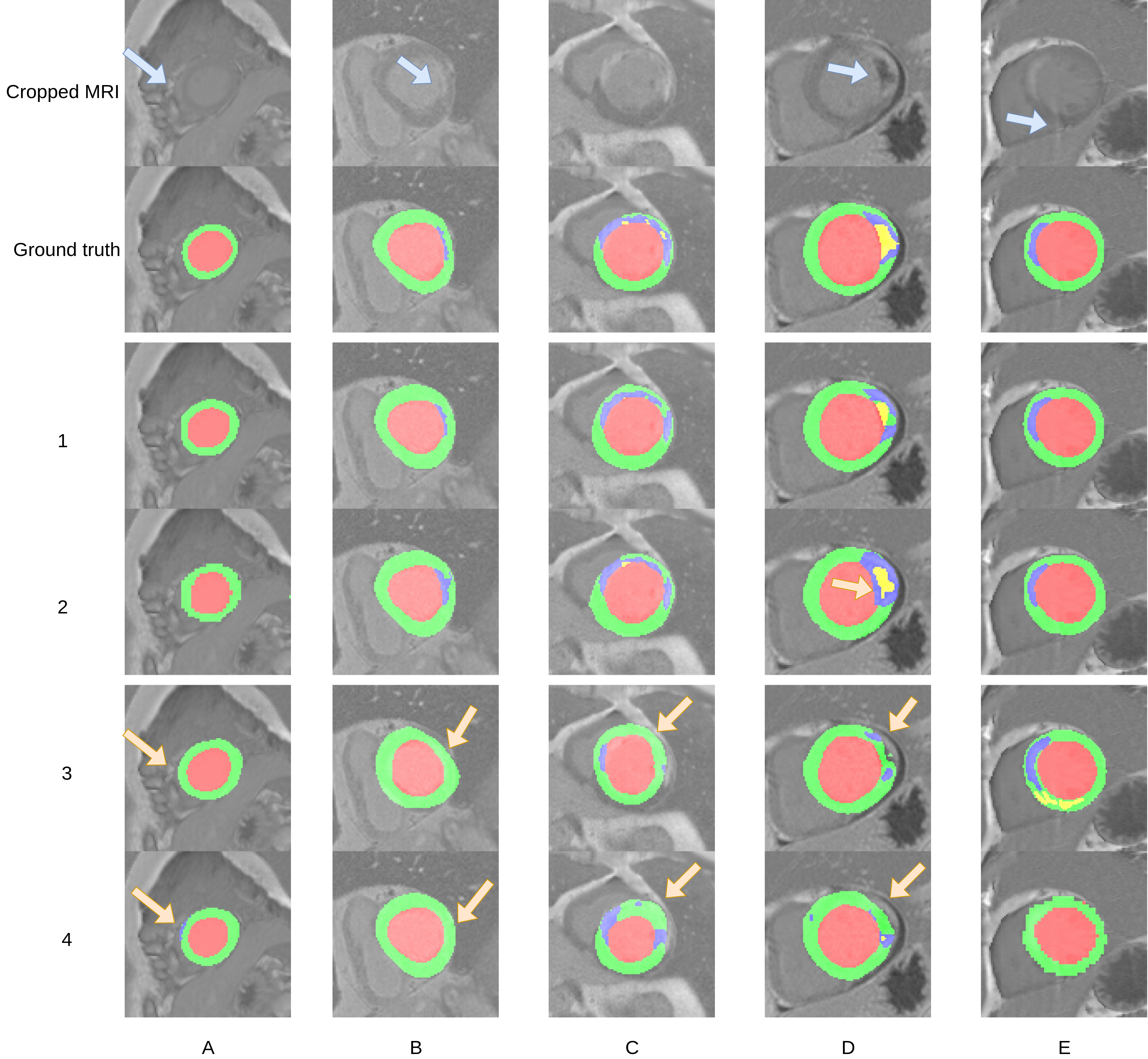}
\caption{Segmentation results on five challenging slices. Rows 1 and 2 denote satisfying segmentation results, rows 3 and 4 denote segmentations to be optimized. Columns A-E denote five slices from different testing cases. The masks in each row may come from different challengers. Blue arrows highlight difficult areas to detect (low contrast, presence of artifact, etc.). Yellow arrows show differences between challengers for specific segments. Cardiac cavity in red, normal myocardium in green, myocardial infarction in blue and PMO in yellow. See details in the text.
}

\label{seg_image}
\end{figure*}

To intuitively present the state-of-the-art segmentation results and the challenges to be overcome, segmentation masks from different challengers on five typical slices are selected. Figure~\ref{seg_image} covers the selected MRI slices and theirs ground truth masks, showing for each slice two well-performed segmentations and two segmentations to optimize. Here are the details:
\begin{enumerate}
    \item Slice A is close to the apex. Therefore only a small part of the right ventricle appears in this slice (blue arrow). Methods on rows 1 and 2 successfully delineated the junction between the left and the right ventricles, while method on row 3 over-estimated the right ventricle and method on row 4 wrongly segmented the right ventricle as a small infarct (yellow arrows).
    \item Slice B involves an infarct that connects the cavity (blue arrow). Methods on rows 3 and 4 failed attributed to the low contrast and narrow width of the infarct. 
    \item Scar tissues in slice C have a broken shape: On the upper side, the scar tissues and the normal myocardium intersperses. The interspersed area was wrongly segmented as normal myocardium on rows 3 and 4 (yellow arrows).
    \item Slice D involves an important PMO area. Although the best adaptive approaches recognized the existence of the PMO, a part of the PMO area was segmented as the normal myocardium (row 1) or the infarct was over-estimated (row 2). Most of the other challengers wrongly segmented the infarct wrapping the PMO as the adipose tissue on the lateral segment of the myocardium (yellow arrows in rows 3 and 4).
    \item Slice E involves an artifact (blue arrow). Reassuringly, for most challengers, the presence of this artifact on the myocardium did not interfere the segmentation while some challengers made atypical errors on this slice. 
\end{enumerate}
In addition, the segmentations of all challengers on one entire exam are provided in the supplementary material.

\subsection{Classification contest}
The classification contest results are listed %together with challengers' key methods 
in Table \ref{res_classification}. The best results were achieved on the merged textual and graphical data. Lourenço  et al., Girum  et al. and Shi  et al. also submitted their classification results relying on sole textual data. The achieved accuracy on the textual data were 70\%, 78\%, and 74\% respectively, which were significantly outperformed by their model with data fusion in Table \ref{res_classification} (82\%, 82\%, and 92\% respectively). The best method failed only on 3 cases among 50, which we can consider as an excellent result.

\section{Discussion}

\subsection{Challenge results}
The overall challenge results were satisfactory. For the segmentation task, the best method obtained a Dice score of 0.879 for the myocardium and of 0.712 for the infarction area. However, compared to the myocardium, scar tissue segmentation still proved to be a daunting task. Methods incorporating complex pipelines or an important amount of parameters did not always show superiority in the results. The best segmentation approach employed two conventional U-Net variants and the configurations of nnU-Net (\cite{nnUnet}) where the first was in 2D and the second was in 3D (\cite{Zhang}). The best pathology classification accuracy is of 92\%. This method employed an encoding CNN to predict the scar volume from the MRI, then concatenated the intermediate volume prediction to other textual features for the final classification. Therefore, it could be assumed that an adaptive approach works more efficiently than attempting heavy networks. The depth of MRI and patient features' semantic information is much less than the data dedicated for human environment applications such as MS COCO and KITTI datasets (\cite{COCO,KITTI}). %This assumption also corresponds to the hypothesis done during the development of the nnU-Net \cite{nnUnet}: 
Unless the appearance of a revolutionary new approach, a better adaptation incorporating the adequate architecture, preprocessing, training and inference {\em etc.} should be a more robust and generalized solution in the domain of medical data.

\begin{table}[htbp!]
\begin{center}
\caption{Results of the classification contest. Best results in bold }
\begin{adjustbox}{width=0.5\textwidth}

\begin{threeparttable}[t]
\begin{tabular}{c|c|c|c|c}
\hline
Challengers &Sensitivity (\%)& Specificity (\%)& Precision (\%)&Accuracy (\%)\\ \hline
Lourenço  et al.  &87.88&70.59&85.29& 82  \\
Ivantsits  et al. &72.73&82.35&88.89& 76 \\
Sharma  et al. &72.73&41.18&70.59& 62 \\
Girum  et al.\tnote{1}  &78.79&88.24&92.86  &82 \\
Shi  et al.\tnote{1} &\textbf{90.91}&\textbf{94.12}&\textbf{96.77}& \textbf{92} \\ \hline
\end{tabular}

\iffalse
\begin{tabular}{c|c|c}
\hline
Challengers & Accu. (\%)&Methods\\ \hline
Lourenço  et al.  & 82& Volume estimation from segmentation, encoded image features \\
Ivantsits  et al. & 78& Segmentation of the myocardium, radiomics features extraction\\
Sharma  et al. & 62& Patient features selection; classifiers boosting, image encoding \\
Girum  et al.\tnote{1}  &84& Patient features selection; volume estimation from segmentation \\
Shi  et al.\tnote{1} & 92&Volume estimation from image encoding \\ \hline
\end{tabular}
\fi
\begin{tablenotes}
\item[1] Co-authors come from the challenge organization team. Do not participate in rankings.
\end{tablenotes}
\end{threeparttable}
\label{res_classification}
\end{adjustbox}
\end{center}
\end{table}

\subsection{Inter-slice correlation}

As discussed by many challengers in the segmentation contest (\cite{Feng,Yang,Zhang}), the inter-slice information is meaningful but tricky. Anatomical facts confirm the correlation between slices, but the cardiac MRI acquisition involves anisotropic voxel size and variable numbers of slices. Such facts require challengers to weigh up the pros and cons about the use of the inter-slice correlation. The winner justified his trade-off (\cite{Zhang}) such as the preliminary segmentation relied only on the intra-slice information and the final segmentation extends the receptive field to the inter-slice information. This approach reproduces the clinical practice: for most of the cases, considering a single slice is sufficient, but to distinguish the infarction and especially the PMO, the neighboring slices should be referred by physicians in case of ambiguity. 

\subsection{Gating and attention mechanism}
The attention mechanism (\cite{attention}) has become a popular topic from serial data as Natural Language Processing (NLP) to computer vision tasks. The attention in neural networks mimics cognitive attention: valuable information should be enhanced and redundant information will be faded out. The attention can be applied to relatively concrete data such as the skip connections (\cite{attentionunet}), or inside a convolutional block for more abstract gating such as SE block and IRB ({\em cf.} \ref{buildingblocks}). Unfortunately, according to the challenge results, the approaches employing the attention mechanism did not prove to outperform the vanilla U-Net or U-Net with conventional building blocks, although some challengers reported its advantage on their split validation set. An ablation study of the attention-based blocks on the state-of-the-art pipeline for the segmentation contest should be worthwhile in future work. 

\subsection{False segmentation and loss functions}
Challengers, especially of the segmentation contest, have taken note on the class imbalance issue. The scar tissues represent a small number of instances in the dataset. The majority of challengers employed basically the weighted cross entropy loss, and optionally the Dice loss or generalized Dice loss (\cite{gdice}). The Dice loss solves the pixel-wise class imbalance problem. However, the vanilla Dice loss does not address the image-wise or the batch-wise imbalance, namely the scar tissues only exist in few images, especially the PMO. Without the weighting, the Dice loss would still suffer from the image-wise imbalance issue: the predictive model would easily assume that such targets do not exist at all, as they do in most batches. It could explain the fact that some challengers under-estimated the scar tissues if they employed the non-weighted multi-class Dice loss, in other words, the generalized Dice loss with equal class weight. 

\subsection{Data variance}
Challengers investigated a variety of data augmentation methods. Such methods have been widely approved for the applications in short of training data. Nevertheless, the generated data should follow the distribution of the original data, thus completely new features should not be produced. According to this hypothesis, data augmentations such as elastic deformation and mix-up should be applied with caution. Overall, the generated features represent a fuzzy concept, only experiments can determinate if the features are bias or not.
Besides the data augmentation, another approach that may increase the data variance of the training data is transfer learning. Some challengers reported the employment of transfer learning with cine MRI from the ACDC dataset (\cite{ACDC}). The cine MRI and the DE-MRI are different acquisition techniques, but both in short axis orientation of the left ventricle. Although the challengers limited the transfer learning on only the myocardium delineation, any approach that may significantly alter the learning characteristics of the model should be undertaken with caution. 

%\IEEEpubidadjcol

\subsection{Clinical implications}
Evaluation of the presence and the extent of the myocardial infarction (with or without PMO) stays crucial in the evaluation of the myocardial viability. The visual estimation by physicians remains the routine approach, although an accurate automatic prediction of the exams as an objective evaluation of the volume and the percentage of diseased myocardium would improve the diagnosis and prognosis steps. Automatic classification allows reducing the time used to do the diagnosis and reduce the inter-expert variability. However, classification software considered as ``black box" must be validated on a large and diverse dataset in order to be accepted in clinical use.  Moreover, the segmentation of the different areas must be done with high accuracy and robustness. Results suggest that automatic myocardial segmentation is now a possible task, but the segmentation of diseased areas needs further development before being integrated into software solutions used in clinical practice. Moreover, in this work, only myocardial infarction is considered, and the proposed approaches must also be tested on other pathologies that involve an abnormal signal in DE-MRI, such as myocarditis or hypertrophic cardiomyopathy.

\section{Conclusion}
DE-MRI is a non-invasive technique providing the assessment of myocardial viability, but it still requires an automatic processing to get objective values of the presence and extent of the disease. In this paper, we have shown that the automatic classification of an exam between normal or pathological is possible. Moreover, the best U-Net based methods provide an accurate delineation of the myocardium. However, the segmentation of the myocardial infarction and particularly that of the PMO area remains challenging, requiring further development to provide the extent of the infarction in a robust manner. These limitations are certainly due to the small size of the disease areas (and then due to the imbalance issue) as the lack of contrast with the surrounding structures.
%The delineation of the myocardium (endocardial and epicardial borders of the left ventricle) as the diseased areas if present (myocardial infarction with or without no-reflow area) allows the accurate calculation of the volume of diseased area (as the percentage of this area compared to the volume of the myocardium). 

\section*{Acknowledgment}

This work was supported by the ADVANCES project founded by ISITE-BFC project (number ANR-15-IDEX-0003) and by the EIPHI Graduate School (contract ANR-17-EURE-0002). 

%\ifCLASSOPTIONcaptionsoff
%  \newpage
%\fi



%\bibliographystyle{IEEEtran}
\bibliographystyle{elsarticle-harv}
\bibliography{elsarticle-template-num.bib}


%\vfill


\end{document}